\documentclass[letterpaper, 10 pt, conference]{ieeeconf}  % Comment this line out if you need a4paper

\IEEEoverridecommandlockouts                              % This command is only needed if 
\usepackage{graphicx} % for pdf, bitmapped graphics files
\usepackage{amsmath} % assumes amsmath package installed
\usepackage{amssymb}  % assumes amsmath package installed
\usepackage{siunitx}
\usepackage{url}
\usepackage{subcaption}
\usepackage{multirow}

\title{\LARGE \bf
Tailoring deep learning for real-time brain-computer interfaces: From offline models to calibration-free online decoding
}

\author{Martin Wimpff$^{1}$, Jan Zerfowski$^{2}$ and Bin Yang$^{1}$% <-this % stops a space
\thanks{$^{1}$Martin Wimpff and Bin Yang are with the Institute of Signal Processing and System Theory,
	University of Stuttgart, 70569 Stuttgart, Germany
	{\tt\small martin.wimpff@iss.uni-stuttgart.de}}
\thanks{$^{2}$Jan Zerfowski is with the Clinical Neurotechnology Laboratory, Department of Psychiatry and Neurosciences, Charit\'e Campus Mitte (CCM), Charit\'e - Universitätsmedizin Berlin, 10117 Berlin, Germany
        }%
}

\begin{document}

\maketitle
\thispagestyle{empty}
\pagestyle{empty}

%%%%%%%%%%%%%%%%%%%%%%%%%%%%%%%%%%%%%%%%%%%%%%%%%%%%%%%%%%%%%%%%%%%%%%%%%%%%%%%%
\begin{abstract}

Despite the growing success of deep learning (DL) in offline brain-computer interfaces (BCIs), its adoption in real-time applications remains limited due to three primary challenges. First, most DL solutions are designed for offline decoding, making the transition to online decoding unclear. Second, the use of sliding windows in online decoding substantially increases computational complexity. Third, DL models typically require large amounts of training data, which are often scarce in BCI applications.\newline
To address these challenges and enable real-time, cross-subject decoding without subject-specific calibration, we introduce real-time adaptive pooling (RAP), a novel parameter-free method. RAP seamlessly modifies the pooling layers of existing offline DL models to meet online decoding requirements. It also reduces computational complexity during training by jointly decoding consecutive sliding windows. To further alleviate data requirements, our method leverages source-free domain adaptation, enabling privacy-preserving adaptation across varying amounts of target data.\newline
Our results demonstrate that RAP provides a robust and efficient framework for real-time BCI applications. It preserves privacy, reduces calibration demands, and supports co-adaptive BCI systems, paving the way for broader adoption of DL in online BCIs. These findings lay a strong foundation for developing user-centered, high-performance BCIs that facilitate immediate feedback and user learning.\newline
\end{abstract}

%%%%%%%%%%%%%%%%%%%%%%%%%%%%%%%%%%%%%%%%%%%%%%%%%%%%%%%%%%%%%%%%%%%%%%%%%%%%%%%%
\section{INTRODUCTION}

\label{sec:introduction}
% what are BCIs
A brain-computer interface (BCI) is a system that measures brain activity and converts it into functionally useful outputs, allowing it to replace, restore, enhance, supplement, or improve the brain’s natural functions \cite{bcisociety}.
% Why BCIs are nice
By facilitating direct communication between the brain and external devices, BCIs can assist individuals with disabilities, enhance cognitive and motor abilities, and improve human-computer interaction \cite{peksa2023state, cervera2018brain}.

% Briefly explain MI
One common approach to controlling a BCI is through the motor imagery (MI) paradigm, in which users imagine moving a body part without physically executing the movement. This mental simulation activates neural mechanisms similar to those involved during actual movement \cite{decety1996neurophysiological}, making MI-BCIs particularly effective for motor rehabilitation in post-stroke patients \cite{cervera2018brain, soekadar2011erd}.
%Specifically, during MI, the power of the $\mu$ and $\beta$ rhythm measured over the sensorimotor area of the brain decreases (event related de-synchronization) and recovers after MI (event related synchronization).

% MI-BCIs are hard to operate
However, especially MI-BCIs are often difficult to operate for inexperienced users as they rely on the endogenous modulation of brain rhythms instead of a neural response to an external stimulus \cite{lee2019eeg}.  
Consequently, a significant portion of users fails to gain control over the system, a phenomenon known as BCI inefficiency or BCI illiteracy \cite{sannelli2019large, zhang2020subject}. 
% possible solutions - Note: DA not mentioned here
In \cite{sannelli2019large} and \cite{zhang2020subject}, recommendations to solve this issue include longer or more effective user training and using advanced decoders to capture complex features.
\newline

User learning is particularly important as there are users that exhibit promising brain modulation during the initial screening but then fail to elicit the proper signals during MI \cite{zhang2020subject}. In other words, among the users who are categorized as inefficient, there are users which have the potential to efficiently control a BCI, which marks a lost opportunity if not addressed appropriately. \newline
Encouragingly, research indicates that BCI usage is a learnable skill \cite{perdikis2018cybathlon, korik2022competing, mcfarland2018brain, shenoy2014combining, orsborn2014closed, sitaram2017closed} where users are able to improve their performance and brain modulation \cite{kober2013learning} through longitudinal training. A crucial part of learning or mastering any skill is the guidance and feedback received during or after the execution \cite{gaume2016psychoengineering}. BCIs providing feedback to the user are referred to as \textit{closed-loop} or \textit{online} BCIs whereas BCIs providing no feedback are termed \textit{open-loop} or \textit{offline} BCIs. Apart from enabling self-regulation and user learning, closed-loop BCIs also increase the attention and motivation of participants during BCI usage \cite{orsborn2014closed, mladenovic2018generic}. Thus, delivering feedback is a fundamental element of BCIs and essential for user learning.  
\newline

The second recommendation provided in \cite{sannelli2019large} and \cite{zhang2020subject} is to use a better decoder, that is able to extract more complex features.
Deep learning models, especially convolutional neural networks (CNNs), are the obvious choice here as they are able to implicitly learn complex distinctive features directly from data and have proven their superiority over traditional decoders for lots of single-trial/offline decoding tasks \cite{craik2019deep}.
Despite these improvements in single-trial decoding, to date, closed-loop systems mostly employ traditional methods such as Common Spatial Patterns (CSP) combined with Linear Discriminant Analysis (LDA) or Support Vector Machine (SVM) classifiers \cite{vavoulis2023review}. This discrepancy between the widespread success of deep learning in single-trial decoding and its rare application in online decoding is remarkable yet remains largely unaddressed in the literature. 
%TODO: move this to M&M
%In \cite{tayeb2019validating} and \cite{jeong2020brain} the general feasibility of deep learning for online control is validated by employing long windows as in single-trial decoding to control a robotic arm. In \cite{tayeb2019validating} the arm was moved at the end of the trial, whereas the approach in \cite{jeong2020brain} is closer to continuous online control as they use sliding windows (\SI{4}{\second} length, \SI{0.5}{\second} shift). While both studies prove that deep learning based decoders are generally useful for control, their settings are not suited for continuous control as the window size is too long and the update frequency too low. \newline 
%Only in \cite{karacsony2019brain, stieger2021benefits, forenzo2024continuous} DL was employed to decode small sliding windows (\SI{0.5}{\second} - \SI{1}{\second}) to continuously control virtual reality feedback or the position of a cursor. 
We hypothesize that there are three primary factors contributing to the limited literature on DL models for closed-loop decoding. \newline

\subsubsection{Unclear transition from offline models}
As DL models are mostly developed to classify whole trials it is unclear how to properly modify them for shorter sequences (e.g., sliding windows). The current options are either to develop an entirely new architecture \cite{karacsony2019brain} or to modify an existing architecture such that it is able to decode shorter windows \cite{stieger2021benefits, forenzo2024continuous}. The first approach is inefficient, as it does not leverage existing architectures and likely requires extensive architectural optimization. While the second approach can build on existing models, the transition from offline to online remains ambiguous and often results in a makeshift solution.

We solve this problem by proposing a new method called real-time adaptive pooling (RAP), which modifies the pooling layers of existing offline deep learning models in a concise manner, parameterized by the requirements of the online decoding task. This approach facilitates a seamless transition from offline to online architectures effectively taking advantage of existing models. 
%We briefly sketched this idea in a previous short conference paper \cite{wimpff2024towards}.

\subsubsection{Computational complexity}
The usage of sliding windows introduces a larger computational demand per trial. For short windows and high update frequencies the number of windows per trial increases. Consequently, the number of forward passes and hence the computational complexity increases. 
While this holds true for any decoder, it particularly affects DL models as the computational complexity of a forward pass is typically higher for DL models than for traditional decoders. Additionally, since DL training is iterative, each window is decoded multiple times during training.

Our novel method RAP can decode consecutive windows of one trial simultaneously while preserving the ability to decode individual windows during inference. RAP achieves this by reusing the intermediate outputs of the network, effectively leveraging the continuous and overlapping nature of sliding windows. 
Through this joint decoding, RAP substantially reduces the computational demand during training, keeping it close to single-trial decoding.

\subsubsection{Data scarcity}
DL models require large amounts of training data, making the within-subject training approach traditionally used for closed-loop decoders impractical. Recording sufficient calibration data for each individual is not only time-consuming but also introduces issues such as user fatigue and reduced data quality due to the non-interactive nature of offline recordings.
These limitations hinder the applicability of DL-based decoders if the conventional within-subject training strategy is followed.

A viable alternative is cross-subject transfer learning, where data from multiple individuals is leveraged to train a generalized decoder. As the number of subjects increases, the individual data requirements decrease. Additionally, cross-subject models can provide immediate feedback, eliminating the need for a subject-specific open-loop calibration phase thus facilitating immediate user learning for unseen users.
However, due to individual variability in EEG patterns, cross-subject models often underperform compared to their within-subject counterparts \cite{lawhern2018eegnet}, necessitating strategies such as \textit{domain adaptation} to mitigate performance degradation.

To address the challenges of deep learning for MI-BCIs, we introduce solutions that bridge the gap between offline and online decoding while improving computational efficiency and data efficiency. We validate our approach on a total of 133 subjects, demonstrating its robustness across a large and diverse user group. Our findings provide a strong foundation for future research and the practical deployment of deep learning in closed-loop BCI systems. Specifically, our contributions are:
\begin{itemize}
	\item We propose real-time adaptive pooling (RAP), a method that seamlessly adapts offline deep learning models for real-time decoding by modifying pooling layers based on the task-specific requirements.
	\item Through RAP, we significantly reduce the computational complexity by jointly processing consecutive sliding windows during training, minimizing redundant forward passes while preserving real-time decoding capabilities during inference.
	\item We systematically investigate transfer learning and domain adaptation techniques to address the data scarcity issue, analyzing their effectiveness under varying levels of data availability. 
	%Specifically, we evaluate supervised few-shot learning, unsupervised domain adaptation (UDA), and online test-time adaptation (OTTA).
\end{itemize}

\section{Materials and Methods}
\begin{figure*}[h]
	\centering
	\includegraphics[width=\textwidth]{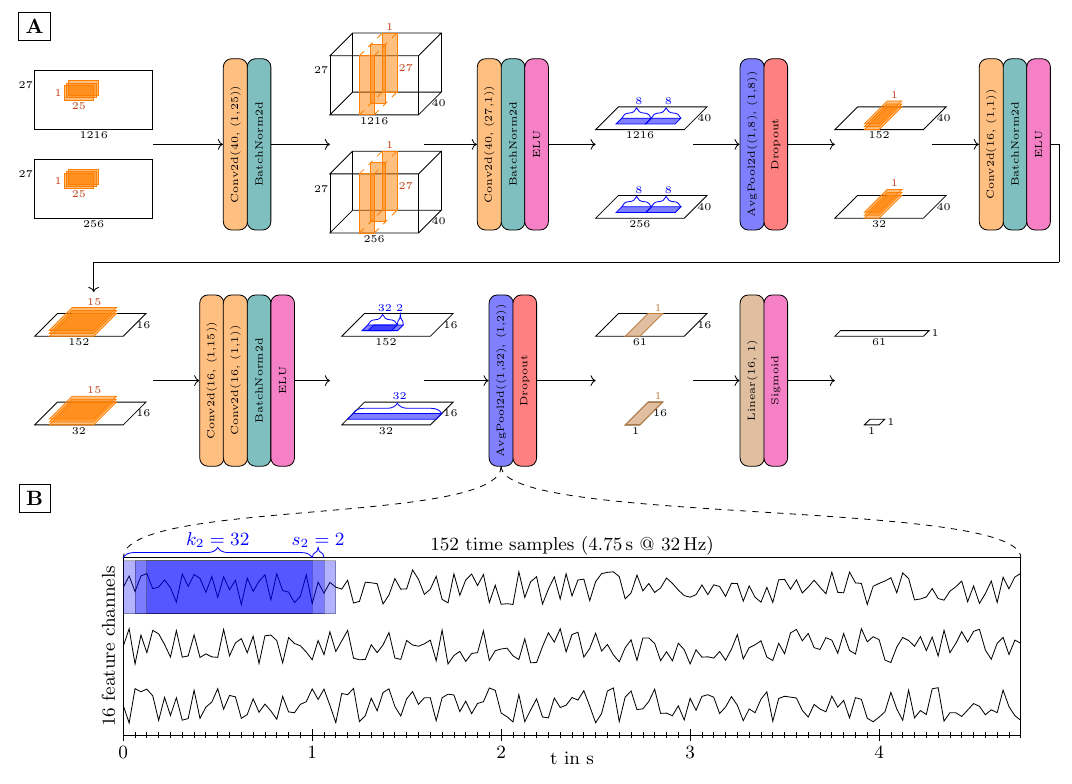}
	\caption{A) BaseNet architecture and output dimensions of joint decoding (upper data stream; trial length $T_t=\SI{4.75}{\second}$) and individual windows (lower data stream; window length $T_w=\SI{1}{\second}$) for the Dreyer2023 dataset.
		Layer names are specified following the PyTorch API conventions. For Conv2d layers, the first value indicates the number of filters, and the tuple represents the kernel size. In the pooling layers, the first tuple indicates the kernel size, while the second tuple specifies the stride. B) Visualization of the sliding window extraction in the second pooling layer of BaseNet for joint decoding.}
	\label{fig:model}
\end{figure*}
\subsection{Real-time adaptive pooling}
\subsubsection{Related work}
Despite the widespread success of deep learning (DL) in single-trial MI decoding, its application in closed-loop BCIs remains limited \cite{vavoulis2023review}. Some studies have explored its feasibility for online control, but with significant limitations. In \cite{tayeb2019validating} and \cite{jeong2020brain}, DL-based decoders were used to control a robotic arm, relying on long windows similar to those used in single-trial decoding. While \cite{tayeb2019validating} moved the arm only at the end of each trial, \cite{jeong2020brain} adopted sliding windows (\SI{4}{\second} length, \SI{0.5}{\second} shift), making it more suited for continuous control. However, both approaches suffer from large window sizes and low update frequencies, making them impractical for real-time applications.

More recent works \cite{karacsony2019brain, stieger2021benefits, forenzo2024continuous} have employed DL to decode shorter sliding windows (\SI{0.5}{\second} – \SI{1}{\second}), enabling continuous control of virtual reality feedback or cursor movement. These studies either developed entirely new architectures \cite{karacsony2019brain} or modified existing models for shorter time lengths \cite{stieger2021benefits, forenzo2024continuous}. However, the former approach is inefficient, requiring extensive architectural optimization, while the latter often results in unclear provisional modifications. Additionally, none of these works address the increased computational complexity introduced by sliding-window decoding.

We argue that the limited adoption of DL for online BCI decoding partly stems from these two key challenges: (1) the lack of a clear methodology to transition from offline to online models and (2) the increased computational burden of processing overlapping sliding windows. To address both issues, we propose real-time adaptive pooling (RAP), a simple yet effective parameter-free method that seamlessly adapts existing offline models for online decoding while maintaining computational efficiency close to single-trial decoding.

\subsubsection{Architecture} 
%Before introducing our method, we briefly outline the rationale behind our choice of architecture for demonstrating our framework.
DL architectures for BCI decoding are highly diverse \cite{craik2019deep}, with significant efforts dedicated to developing increasingly complex models \cite{sartzetaki2023beyond}. However, research has shown that simple and shallow CNNs achieve comparable performance across different datasets while maintaining lower computational complexity \cite{wimpff2024eeg, sartzetaki2023beyond, forenzo2024continuous}. Given this, we employ BaseNet \cite{wimpff2024eeg}, a modern evolution of the widely used shallow architectures EEGNet \cite{lawhern2018eegnet} and ShallowNet \cite{schirrmeister2017deep}, as the foundation for showcasing our framework.\newline
The main contributions of this paper are the proposed real-time adaptive pooling (RAP) method, along with various domain adaptation techniques, to enable cross-subject generalization for online decoding. These ideas are applicable to any convolutional architecture that employs pooling layers.\newline

\subsubsection{RAP}
RAP provides a parameter-free approach for seamlessly transitioning offline deep learning models to online decoding by re-parameterizing the pooling layers based on the specific requirements of the online decoding problem. By adjusting the kernel sizes $k_i$ and stride lengths $s_i$ according to the update frequency $f_u$ and window length $T_w$, RAP enables any CNN to efficiently process sliding windows. This structured adaptation ensures that models originally designed for offline use can be directly applied in real-time scenarios without requiring additional architectural modifications.

Online decoding requires a high update frequency (e.g., \SI{16}{\hertz}) to provide continuous feedback, leading to a significant overlap between consecutive windows (e.g., $\frac{15}{16}$ for a \SI{1}{\second} window). This redundancy is also present in the intermediate layers of deep learning models, as previously noted in \cite{schirrmeister2017deep}. 
While \cite{schirrmeister2017deep} introduced \textit{cropped training} to stabilize single-trial decoding, RAP takes this idea further by systematically exploiting the overlap to reduce computational complexity. By jointly decoding all windows of one trial during training, RAP minimizes redundant forward passes, significantly improving efficiency. At the same time, the structured pooling modifications ensure that each window remains independently decodable during inference, maintaining real-time performance without additional computational overhead during training.

Since our method is applicable to any CNN, we will explain our approach for the general case of $P$ pooling layers. \newline
The first $P-1$ pooling layers are used to downsample the original input from a sampling frequency $f_s$ to an intermediate frequency $f_{\text{inter}}$.
\begin{equation}
	f_{\text{inter}} = \frac{f_s}{\prod_{i=1}^{P-1}k_i}, \quad k_i=s_i \forall i\leq P-1
\end{equation}
The values of $k_i$ and $s_i$ can be chosen arbitrarily as long as the resulting intermediate frequency $f_{\text{inter}}$ is equal to the update frequency $f_u$ or an integer multiple of it. For BaseNet ($P=2$), we use $f_s=\SI{256}{\hertz}, k_1=s_1=8$ and consequently $f_{\text{inter}}=\SI{32}{\hertz}$ (compare Fig. \ref{fig:model}A). 
For models with only one pooling layer ($P=1$), the downsampling stage would be dropped and $f_{\text{inter}}=f_s$.\newline
The final pooling layer $P$ is used to extract overlapping sliding windows which fulfill the requirements of the online application (window length $T_w$ and update frequency $f_u$). Specifically, the kernel size $k_P$ is chosen based on the window length $T_w$ whereas the stride $s_P$ depends on the update frequency $f_u$.
\begin{equation}
	k_P = f_{\text{inter}}\cdot T_w, \quad s_P = \frac{f_{\text{inter}}}{f_u}
\end{equation}
For our datasets and decoding problem $T_w=\SI{1}{\second}$ and $f_u=\SI{16}{\hertz}$. Hence $k_2=32$ and $s_2=2$.
The extraction of the overlapping sliding windows during joint decoding via the last pooling layer is visualized in Fig. \ref{fig:model}B for the Dreyer2023 dataset.

\begin{figure}
	\centering
	\includegraphics[width=\columnwidth]{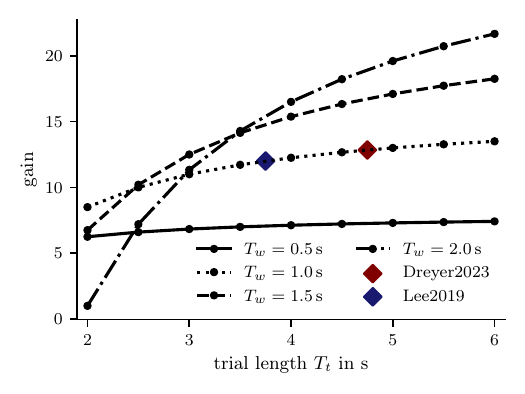}
	\caption{Computational gain of joint decoding for $f_u=\SI{16}{\hertz}$ and different trial lengths $T_t$ and window lengths $T_w$.}
	\label{fig:gain}
\end{figure}

A key advantage of RAP is its ability to significantly reduce computational complexity during training. Instead of processing each sliding window separately, RAP allows the model to jointly decode multiple overlapping windows, leveraging their redundancy to minimize redundant forward passes.
For any input $X$ with time length $T=T_w+(N_w-1)\cdot\frac{1}{f_u}$, RAP ensures that the model outputs a prediction vector of length $N_w$, irrespective of the number of jointly processed windows:
\begin{equation}
	X\in\mathbb{R}^{C\times T\cdot f_s}; \quad y = f(X) \in \mathbb{R}^{N_w}.%; \quad N_w = (T-T_w)\cdot f_u + 1
\end{equation}
Decoding all windows of one trial of length $T_t$ jointly requires the calculation of $T_t\cdot f_s$ samples, whereas decoding each of the $N_w$ windows individually requires the calculation of $N_w\cdot T_w\cdot f_s$ samples.
The reduction in computational complexity during training is thus
\begin{equation}
	 \frac{T_w}{T_t}\cdot N_w. 
\end{equation}
For our datasets ($T_t=\SI{4.75}{\second}/\SI{3.75}{\second}$, $T_w=\SI{1}{\second}$, $f_u=\SI{16}{\hertz}$), this results in computational savings by factors of $\sim12.84$ and $12$, respectively. The computational gain of joint decoding is visualized in Fig. \ref{fig:gain} for different trial and window lengths. Table \ref{tab:trial-window} additionally provides an overview of all dataset and RAP hyperparamters with units.

Importantly, RAP ensures that each window remains independently decodable during inference, even when joint decoding is used during training. The structured pooling modifications preserve the independence of sliding windows as much as possible. While slight differences may occur between a window decoded jointly with others and the same window decoded individually, due to the padding properties of the convolutional layers, these variations are negligible since inference is always performed on single windows.

\begin{table}
	\centering
	\caption{Dataset and RAP hyperparameters with units for both datasets.}
	\label{tab:trial-window}
	\resizebox{\columnwidth}{!}{
		\begin{tabular}{|c|c|c|c|c|c|c|c|c|}
			\hline
			$\text{dataset}$ & $T_t$ & $T_w$ & $f_s$ & $f_{\text{inter}}$ & $f_u$ & $k_i$ & $s_i$ & $N_w$ \\ \hline
			Dreyer2023   & \SI{4.75}{\second}  & \SI{1}{\second}  & \SI{256}{\hertz} & \SI{32}{\hertz} & \SI{16}{\hertz}  & [8, 32]  & [8, 2]  & 61    \\ \hline
			Lee2019   & \SI{3.75}{\second}  & \SI{1}{\second}  & \SI{256}{\hertz} & \SI{32}{\hertz} & \SI{16}{\hertz}  & [8, 32]  & [8, 2]   & 45   \\ \hline
		\end{tabular}
	}
\end{table}

During inference, the trained DL model processes each individual window of time length $T_w=\SI{1}{\second}$ independently, as required for real-time online decoding (see second data stream of Fig. \ref{fig:model}A). This is crucial for maintaining low latency in the BCI. Unlike joint decoding used during training, which accelerates processing, applying it during inference would require waiting until the end of the trial \cite{tayeb2019validating} to generate a prediction vector, which would violate the real-time constraints. To meet the real-time constraint, each window must be decoded within $\frac{1}{f_u}=\SI{62.5}{\milli\second}$. The inference times will be reported later in section \ref{subsec:res-inference}.
\begin{figure*}
	\centering
	\includegraphics[width=\textwidth]{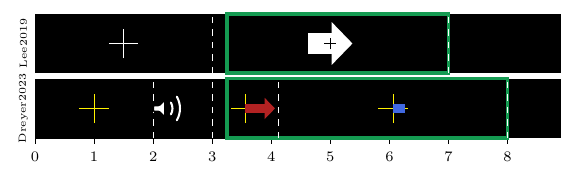}
	\caption{Trial structure for both datasets. Each trial starts with a fixation cross. For the Dreyer2023 dataset this is followed by an auditory signal. The cue occurs after \SI{3}{\second} for both datasets. For the Dreyer2023 dataset the cue is present for \SI{1.25}{\second} seconds and followed by a \SI{3.75}{\second} feedback phase, with the feedback indicated by a blue bar. For the Lee2029 dataset, the cue is present during the whole trial and the feedback is indicated by the position of a small cross within the cue arrow. Between the trials a blank screen is displayed. The time period used for classification is indicated by the green outline.}
	\label{fig:trial}
\end{figure*}
\subsection{Datasets}
We employ the large EEG Database \cite{dreyer2023large} published by \textit{Dreyer et al.} in 2023 and the OpenBMI dataset \cite{lee2019eeg} published by \textit{Lee et al.} in 2019. In this section we will briefly present both datasets.\newline
The Dreyer2023 dataset contains motor imagery EEG data from 87 subjects of which we exclude 8 subjects due to artifacts and missing data yielding effectively 79 subjects. The participants were asked to imagine movements of the right or left hand during the trials. The EEG montage consists of 27 scalp electrodes distributed over the motor cortex. \newline
For each subject a total of 240 trials were recorded. The trials were recorded in 6 runs with 40 trials per run. During the first two runs, sham feedback was provided to familiarize the user with the visual feedback during the closed-loop phase. With the trials from the first two runs, a subject-specific decoder consisting of CSP and LDA was trained. This algorithm delivered the feedback during the last 4 runs. The feedback was updated every \SI{62.5}{\milli\second}, i.e., with an update frequency of \SI{16}{\hertz}.\newline
Each trial lasts a total of 8 seconds and the trial structure is visualized in Figure \ref{fig:trial}. After an initial fixation cross and an auditory signal, the cue is presented for 1.25 seconds followed by 3.75 seconds of feedback. Since \cite{dreyer2023large} used a sliding window of one second and provided feedback from \SI{4.25}{\second} after trial onset, we use the data from \SI{3.25}{\second} - \SI{8}{\second} after trial onset (green outline in Fig. \ref{fig:trial}) for classification to match their setting. This yields an effective trial length of 4.75 seconds. The update  frequency of \SI{16}{\hertz} together with a window length of \SI{1}{\second} lead to a $\frac{15}{16}=93.75\%$ overlap between consecutive windows and a total of $(\SI{4.75}{\second}-\SI{1}{\second})\cdot\SI{16}{\hertz}+1=61$ windows per trial.\newline
For preprocessing, we employ a \SI{5}{\hertz} - \SI{35}{\hertz} bandpass filter as  in \cite{dreyer2023large} for all subjects and downsample the data from \SI{512}{\hertz} to \SI{256}{\hertz}.

The Lee2019 dataset contains EEG data from 54 subjects performing the same binary classification task as in \cite{dreyer2023large} recorded with 62 electrodes of which the 20 sensors above the motor cortex were selected. The dataset is divided into an offline and an online session, each containing a total of 100 trials per subject. Similarly to the Dreyer2023 dataset, a combination of CSP and LDA was used to provide the visual feedback during the closed-loop phase. No (sham) feedback was provided during the offline session.\newline
Each trial lasts a total of 7 seconds, with 3 seconds of fixation followed by 4 seconds of MI. The cue was present during the whole 4 seconds of imagery and the feedback was indicated by the position of a small cross within the arrow used for the cue. As \textit{Lee et al.} originally uses a relatively slow online setting (\SI{1.5}{\second} window, \SI{2}{\hertz} update frequency) and further cropped their trials to a length of \SI{2.5}{\second}, we decided to take over the setting from \textit{Dreyer et al.} to have a more challenging task as well as better comparability between the datasets. Starting from \SI{0.25}{\second} after the cue onset, this yields a trial length of \SI{3.75}{\second} (green outline in Fig. \ref{fig:trial}) and results in 45 windows per trial.\newline
For preprocessing, we employ a \SI{8}{\hertz} - \SI{30}{\hertz} bandpass filter as in \cite{lee2019eeg} for all subjects and downsample the data from \SI{1000}{\hertz} to \SI{256}{\hertz}.

\subsection{Source training}
\subsubsection{Data split}
Generally speaking, there are two main settings to train BCI models, one is the within-subject setting, the other one is the cross-subject setting. 
While we will focus on the cross-subject setting due to the data requirements of DL mentioned earlier, we still include the within-subject setting for comparison and completeness.\newline
As both datasets have a slightly different structure, we will use the terms offline (first two runs in the Dreyer2023 dataset, first session in the Lee2019 dataset) and online data (last four runs in the Dreyer2023 dataset, second session in the Lee2019 dataset) in the following to explain the data splits.

\textit{Within-subject}: We train one model per subject on its offline data and test the model on the online data of the same subject. We repeat this process for every subject and report the average and standard deviation between the subjects.

\textit{Cross-subject}: We train our models in a leave-one-subject-out cross-subject setting. This means that we train one model per subject using the offline data of all remaining subjects. In the domain adaptation context we refer to these subjects used for training as \textit{source} subjects. The subject used for testing is called the \textit{target} subject. Independent of the setting used during subsequent domain adaptation, we always report the test accuracy on the online data of the target subject which is not used at all during training. We repeat this process for every subject and report the average and standard deviation between the subjects. \newline
Our cross-subject data split resembles a realistic scenario, where a small amount of data from a large number of subjects is available at the beginning of a study. Afterwards, the online BCI usage, user training and model adaptation can be highly individual.
\subsubsection{Training procedure}
We train all models with the original training procedure of BaseNet described in \cite{wimpff2024eeg} for both data splits. Specifically, we use an Adam optimizer with a learning rate of $10^{-3}$ and train each model for 100 epochs using a learning rate scheduler with 20 warmup epochs and a subsequent cosine decay. As the training process is stochastic (e.g., subject selection, data shuffling, weight initialization and dropout), we train each model for five different random seeds and report the average of these five runs. The complete source code is available at \url{https://github.com/martinwimpff/eeg-online}.
\subsection{Transfer learning}
Transfer learning (TL) typically involves utilizing knowledge or data from a \textit{source} domain to solve a task in the \textit{target} domain. This approach reduces the amount of target data needed to address the target task \cite{wu2022transfer,ko2021survey}.

Formally, a domain comprises a feature space $\mathcal{X}$ and a corresponding marginal probability distribution $P(X)$ with $X\in\mathcal{X}$. A task includes a label space $\mathcal{Y}$, a corresponding marginal probability distribution $P(y)$ with $y\in\mathcal{Y}$, a conditional probability distribution $P(y|X)$ and a prediction function $f(X)$ \cite{wu2022transfer}.\newline
This definition opens up many possible TL settings, e.g., cross-dataset settings where the feature spaces of two domains \cite{han2023eeg, gu2023generalizable} or the label spaces \cite{he2020different} can be different. However, in this work we consider the case where the acquisition setup as well as the task is consistent across subjects, hence they share a common feature and label space. What differs between the subjects is their marginal distribution $P(X)$ as well as the conditional distribution $P(y|X)$. This is considered the most common TL setting in BCI decoding \cite{wu2022transfer}.

In our setting, the source domain consists of $N-1$ source subjects with $N^\text{off}_t$ labeled offline trials $\{[(X^\text{off}_{s_i,j}, y^\text{off}_{s_i,j})]_{j=1}^{N^\text{off}_t}\}_{i=1}^{N-1}$ per subject. The target domain that the model is evaluated on are the $N^\text{on}_t$ online trials of the target subject $\{(X^\text{on}_{t_N,j}, y^\text{on}_{t_N,j})\}_{j=1}^{N^\text{on}_{t}}$. \newline
The initially trained cross-subject decoder has to have a decent generalization capability to immediately provide good feedback to unseen target users. This starting point, i.e., a good domain generalization \cite{kostas2020thinker}, is especially important to help subjects who initially have problems to elicit the proper brain signals \cite{kumar2024transfer,sultana2023towards}.

To improve the source decoder as target data becomes available, we additionally examine different levels of domain adaptation which are explained in the following section. Importantly, to ensure the broader applicability of our approach, we impose a privacy constraint on the source subjects by adopting a source-free domain adaptation framework, where source data is not accessible during the adaptation process \cite{ju2020federated, wang2023privacy}. 

\subsection{Domain adaptation}
Since there is a large number of different approaches in DA, we will provide a simplified but clear categorization of the most common source-free DA settings from a data availability perspective. 
We distinguish between labeled calibration data, unlabeled calibration data and no calibration data, which yields the categories supervised DA, unsupervised DA and online test-time adaptation (OTTA), respectively. This distinction makes sense for motor imagery as subject-specific calibration data is not always available. If calibration data was previously recorded, it can be labeled or unlabeled, e.g., in the case of voluntary imagined movements. In the following sections we will briefly describe the different scenarios we examine, how the datasets are used to investigate them, and provide a quick overview of the relevant literature.\newline
Afterwards, we will introduce the domain adaptation methods we used and how they are implemented under the different settings.
\subsubsection{Supervised domain adaptation}
Both datasets contain offline and online data, with the latter used for testing as stated previously.
The remaining offline target data $\{(X^\text{off}_{t_N,j}, y^\text{off}_{t_N,j})\}_{j=1}^{N^\text{off}_{t}}$ is the subject-specific calibration data. In this setting, we use the calibration data to fine-tune the model towards the target subject in a supervised fashion. This procedure is also termed supervised few-shot learning or supervised fine-tuning/calibration and is the most common domain adaptation method for BCI decoding \cite{ko2021survey}.
\cite{han2023eeg, guetschel2023transfer, xie2023cross} use a variety of datasets to train their models, followed by fine-tuning on either the target dataset \cite{han2023eeg} or on data from the target subject \cite{guetschel2023transfer, xie2023cross}. In a manner similar to our approach, the studies \cite{sartzetaki2023beyond, ouahidi2023strong, junqueira2024systematic, duan2021ultra} perform cross-subject training and subsequently fine-tune on the target subject’s data, while maintaining a source-free setting. By contrast, other approaches \cite{he2020different, xu2021selective, an2020few, liu2022subject} leverage both source and target data in combination to adapt their models. For instance, \cite{he2020different} aligns source and target labels using a few labeled target samples, while \cite{liu2022subject} performs unsupervised joint alignment of source and target features followed by supervised fine-tuning. Additionally, \cite{xu2021selective} and \cite{an2020few} train decoders using both source and target data in a supervised manner.
\subsubsection{Unsupervised domain adaptation} 
The data split remains the same as in the previous setting, but with the key difference that the labels of the subject-specific data are not available ($\{(X^\text{off}_{t_N,j})\}_{j=1}^{N^\text{off}_{t}}$). This setting is sometimes also called unsupervised few-shot learning or unsupervised offline fine-tuning/calibration and has also already been widely explored in the literature. 
A common strategy in unsupervised domain adaptation (DA) is to calculate target-specific statistics to normalize or align the raw target data \cite{ouahidi2023strong, ouahidi2024unsupervised, xu2021improving, zoumpourlis2024motor}. More advanced methods \cite{gu2023generalizable, bakas2023latent} further adapt the model's batch normalization layers to standardize the data between layers. \cite{xia2022privacy} employed an unsupervised loss to reduce uncertainty and enforce consistency regularization, adapting multiple layers within the model.
Additionally, some unsupervised DA approaches involve joint use of source and target data. For example, \cite{xu2021selective} selects source subjects for training based on unsupervised target samples, while other methods aim to reduce the distance between source and target distributions by re-centering them \cite{zhuo2024riemannian} or by minimizing a discrepancy-based loss function \cite{jimenez2020custom, zhang2020manifold}.
\subsubsection{Online test-time adaptation} This setting is calibration-free and therefore only exploits the incoming stream of unlabeled test data (the online data of the target subject) for adaptation.
Importantly, this stream of data means that any adaptation method receives one sample at a time (single-instance OTTA) and can only use the current sample and the previously decoded samples for adaptation.
%Sometimes, an adaption buffer is built online to store the previous samples. 
The OTTA setting is relatively unexplored in the BCI context with \cite{mao2023online} performing seizure prediction and only \cite{li2023t, kumar2024transfer, wimpff2024calibration} performing motor imagery decoding. In \cite{kumar2024transfer} traditional methods are used whereas in \cite{li2023t, wimpff2024calibration} deep learning is employed. 
\subsubsection{Alignment}
Alignment is among the easiest and hence most common approaches to mitigate distribution shifts in EEG data and there are basically two variants of alignment, Euclidean alignment (EA) \cite{he2019transfer} and Riemannian alignment (RA) \cite{zanini2017transfer}. 
%Alignment generally modifies the input $X$ of an algorithm which enables it to be used for any decoding algorithm.
The idea of alignment is to compute a reference state per domain and then to re-center the data from each domain based on this reference state.
The underlying assumption is that the re-centered brain activity between domains is similar, i.e., the difference between domains lies (predominantly) in the reference state \cite{zanini2017transfer}.\newline
A reference covariance matrix is computed by either taking the Euclidean mean (arithmetic mean) or the Riemannian mean (geometric mean) of all covariance matrices of the input windows $X_{s_i,j}\in\mathbb{R}^{C\times T_w\cdot f_s}$ in one domain $i$ with $\delta$ being the Riemannian distance.
\begin{align}
	\bar{R}_i^{EA} &= \frac{1}{N_t\cdot N_w}\sum_{j=1}^{N_t\cdot N_w}X_{s_i,j}\cdot X_{s_i,j}^T \\
	\bar{R}_i^{RA} &= \arg\min_R\sum_{j=1}^{N_t\cdot N_w}\delta^2(R,X_{s_i,j}\cdot X_{s_i,j}^T)
\end{align}
The major disadvantage of the Riemannian mean $\bar{R}_i^{RA}$ is that it has no closed form solution and hence the mean has to be computed iteratively. The advantage of RA is that the geometric mean is less susceptible to outliers compared to the arithmetic mean.\newline
The alignment of the trials is similar between EA and RA.
\begin{equation}
	%\tilde{X}_{s_i,j}^{EA} = \bar{R}_i^{EA,-1/2}\cdot X_{s_i,j};\qquad
	%\tilde{X}_{s_i,j}^{RA} = \bar{R}_i^{RA,-1/2}\cdot X_{s_i,j} 
	\tilde{X}_{s_i,j} = \bar{R}_i^{-1/2}\cdot X_{s_i,j}
	%\frac{1}{N_s}\sum_{j=1}^{N_s}\tilde{X}_{s_i,j}^{EA}\cdot \tilde{X}_{s_i,j}^{EA,T}=\bar{R}_i^{EA,-1/2}\bar{R}_i^{EA}\bar{R}_i^{EA,-1/2}=I
\end{equation}
After alignment, the aligned windows $\tilde{X}_{s_i,j}$ are then used for decoding instead of the unaligned windows $X_{s_i,j}$. \newline
One remaining issue of alignment is the need of a reference state for each new domain. Consequently, unlabeled calibration data is necessary. For the supervised and unsupervised setting we use the offline data of the target subject to calculate the reference matrix. For the online setting there is no calibration data available and thus the reference matrix has to be estimated online with the incoming stream of target data as done in \cite{he2019transfer, zanini2017transfer, xu2020cross, li2023t, wimpff2024calibration, kumar2024transfer}. These approaches mainly differ in how they weight the incoming samples (e.g., equal \cite{xu2020cross, li2023t, kumar2024transfer}, linear \cite{zanini2017transfer} or exponential \cite{wimpff2024calibration}) and how many samples they use to compute the reference state, i.e., using an adaptation buffer \cite{he2019transfer, wimpff2024calibration} or not.\newline 
As our primary benchmarking method \cite{kumar2024transfer} uses equal weighting and no adaptation buffer for their online setting, we will match this in our online setting for comparability.
\subsubsection{Adaptive Batch Normalization}
Adaptive Batch Normalization (AdaBN) \cite{li2016revisiting} is a simple yet very effective DA strategy that changes the statistics in all Batch Normalization (BN) layers of a model to adapt to a new domain.\newline
Generally, BN addresses the internal covariate shifts during training by normalizing the data within the model to speed up convergence and stabilize training. During training, each BN layer normalizes each batch of data by using the batch statistics. Additionally, each BN layer keeps an exponential moving average of the training statistics (i.e., mean $\mu_S$ and variance $\sigma_S^2$).
During inference, these training statistics are then used to normalize the test samples.\newline
However, if there is a distribution shift between training and inference, this normalization fails, i.e., the data is no longer normalized to zero mean and unit variance and hence the performance drops. AdaBN solves this problem by replacing the source statistics $\mu_S$ and $\sigma_S^2$ by target statistics $\mu_T$ and $\sigma_T^2$. As with alignment, unlabeled target data is necessary for AdaBN to collect the target BN statistics.
For the supervised fine-tuning setting, AdaBN is not necessary, as the internal statistics in the BN layers get updated implicitly during supervised fine-tuning with the offline target data. 
For the unsupervised setting we calculate the statistics using the offline data of the target subject. 
% as done in previous works \cite{jimenez2020custom, gu2023generalizable, bakas2023latent, xu2021improving, ouahidi2023strong}.
% \cite{ouahidi2024unsupervised,li2023t, wimpff2024calibration, mao2023online, wang2023privacy, schneider2020improving}
For the online setting, we will update the initial source statistics after every window $X$ from the online target data using a small momentum $\alpha=0.001$ \cite{schneider2020improving}.
\begin{align}
	\mu_i &= (1-\alpha)\cdot\mu_{i-1} + \alpha\cdot\text{E}[X]; \qquad \mu_0=\mu_S\\
	\sigma_i^2&=(1-\alpha)\cdot\sigma_{i-1}^2 + \alpha\cdot\text{Var}[X]; \qquad \sigma_0^2=\sigma_S^2	
\end{align}

\subsection{Benchmark method}
We selected the adaptive Riemannian framework from \textit{Kumar et al.} \cite{kumar2024transfer} as our primary benchmark as it is a very recent publication that employs traditional machine learning, is adaptive in different data settings (both unsupervised and supervised) and was originally presented in a very similar setting to ours. This makes it a competitive and fair choice to evaluate our approach against the current state-of-the-art in online BCI-MI decoding.\newline
Specifically, \textit{Kumar et al.} use a minimum distance to mean (MDM) Riemannian geometry classifier that uses RA by default, always requiring unlabeled target data, thereby categorizing it as unsupervised domain adaptation.
Additionally, \textit{Kumar et al.} introduced an online RA estimator termed Generic Recentering (GR) and a supervised fine-tuning strategy named Personally Adjusted Recentering (PAR). This results in one method for each calibration data setting: RiemannMDM+PAR for supervised fine-tuning, RiemannMDM for unsupervised fine-tuning, and RiemannMDM+GR for online adaptation.

\subsection{Evaluation}
Our most important metric is the trial-wise accuracy (TAcc) from \cite{dreyer2023large}. We average the prediction probabilities of all windows of each trial and check whether the mean prediction equals the correct label. Through the averaging, more confident predictions are favored \cite{dreyer2023large}. Additionally, we also provide the trial-wise accuracy without averaging which we call unaveraged trial-wise accuracy (uTAcc). Further, we provide the window-wise accuracy (WAcc), i.e., the percentage of windows classified correctly.
To compare our methods, we perform one-sided paired t-tests between the trial-wise accuracies.

\section{Results}

\subsection{Within-subject}
\begin{figure}
	\centering
	\includegraphics[]{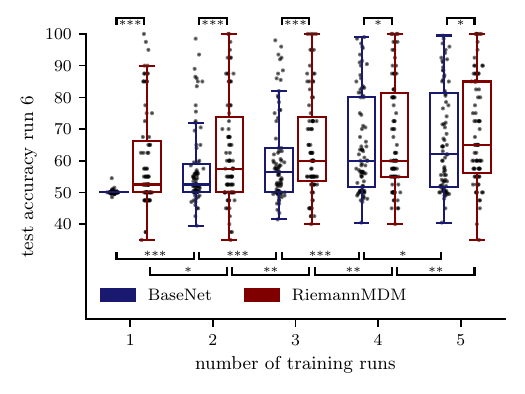}
	\caption{Within-subject results for Dreyer2023 and different amounts of training data. Each dot resembles a subject, the stars within the brackets resemble the different significance levels ($p<0.05$ (*), $p<0.01$ (**) and $p<0.001$ (***)) when comparing two experiments connected by that bracket.}
	\label{fig:results-within}
\end{figure}

The results of the within-subject experiments are displayed in Table \ref{tab:results-within}. Two main observations can be made for the within-subject setting. First, RiemannMDM outperforms our method for both datasets with $p<0.001$ for the Dreyer2023 dataset and $p=0.065$ for the Lee2019 dataset. Second, although RiemannMDM achieves higher average accuracy, it also exhibits greater variance across subjects. \newline
Given the significant performance gap for the Dreyer2023 dataset, we conducted an additional ablation study to assess the impact of training data quantity on performance (see Fig. \ref{fig:results-within}). We varied the number of runs used for training and evaluated the performance using data from the final run. While both decoders generally improved with increased training data, BaseNet demonstrated a larger performance gain than RiemannMDM, leading to a narrowing performance gap between both approaches as more data became available.

\subsection{Cross-subject}

\subsubsection{Amount and diversity of training data}
\begin{figure}
	\begin{subfigure}{0.5\textwidth}
		\centering
		\includegraphics[width=\columnwidth]{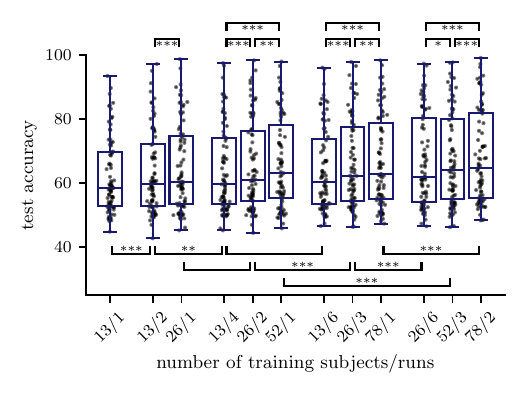}
		\caption{Dreyer2023 dataset}
		\label{fig:results-diversity-large}
	\end{subfigure}
	\hfill
	\begin{subfigure}{0.5\textwidth}
		\centering
		\includegraphics[width=\columnwidth]{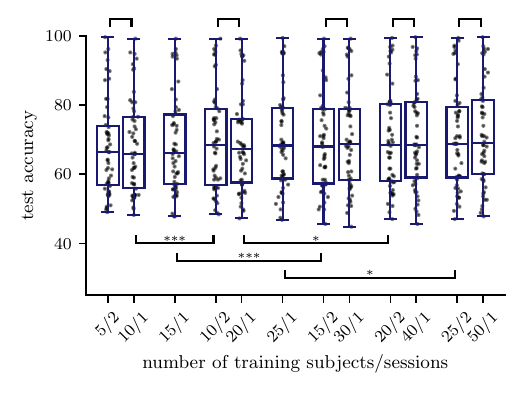}
		\caption{Lee2019 dataset }
		\label{fig:results-diversity-openbmi}
	\end{subfigure}
	\caption{Cross-subject results for different data compositions. Each dot resembles a subject, the stars within the brackets resemble the different significance levels ($p<0.05$ (*), $p<0.01$ (**) and $p<0.001$ (***)) when comparing two experiments connected by that bracket. Empty brackets resemble no statistical significance between two experiments.}
	\label{fig:results-diversity}
\end{figure}

Evaluating the source model's performance across different data compositions is crucial before applying additional domain adaptation. This assessment helps determine the model’s domain generalization capabilities and the quality of feedback an unseen user can receive without any user-specific calibration. Additionally, it provides insights into the data requirements for developing effective cross-subject models, guiding the data acquisition for optimal performance.\newline
To investigate the behavior of our method for the cross-subject setting under different data compositions, we trained BaseNet for different amounts of training subjects (randomly picked) and runs/sessions (sequentially picked, e.g., 4 runs equals picking the first four runs) and present the results in Fig. \ref{fig:results-diversity}. We performed statistical tests between compositions with the same total amount of training data but different amounts of training subjects (above the boxplots) as well as between different amounts of training data using the same number of subjects (below the boxplots).
For both datasets, increasing the number of runs/sessions per subject improves the final performance. The impact of subject diversity yields two different outcomes. In the Dreyer2023 dataset, given the same amount of training data, increasing the number of subjects always produces significantly better results compared to increasing the number of runs per subject. Conversely, for the Lee2019 dataset, this pattern does not hold. Performance remains quite similar regardless of the composition, and for some compositions, increasing the number of sessions per subject even leads to a slightly better performance. 
\begin{table}
	\centering
	\caption{Within-subject results. Results above the double line are for the Dreyer2023 dataset, below are for the Lee2019 dataset.}
	\label{tab:results-within}
	\begin{tabular}{ |p{2.8cm}||p{1.3cm}|p{1.3cm}|p{1.3cm}|  }
		\hline
		Method  & TAcc(\%) & uTacc(\%) & WAcc(\%)\\
		\hline
		RiemannMDM &$63.1\pm14.8$&$62.6\pm14.5$&$60.0\pm11.5$\\ 
		BaseNet&$58.0\pm11.9$&$57.5\pm11.5$&$56.1\pm8.9$\\
		\hline \hline
		RiemannMDM &$62.5\pm15.2$&$62.1\pm14.8$&$60.5\pm13.1$\\ 
		BaseNet&$60.4\pm12.7$&$59.5\pm12.5$&$58.6\pm10.9$\\
		\hline
	\end{tabular}
\end{table}

\subsubsection{Domain adaptation}
The cross-subject results under different DA settings are shown in Table \ref{tab:results-large} and \ref{tab:results-openbmi} for the Dreyer2023 dataset and the Lee2019 dataset, respectively. Based on the different DA settings, we separate our results in three sections and compare them to the corresponding benchmark method by \textit{Kumar et al.}\newline
For the supervised fine-tuning and the unsupervised fine-tuning setting, BaseNet performs significantly better than the corresponding benchmark approach regardless of the dataset and the specific DA method. However, for the Lee2019 dataset, using supervised or unsupervised DA deteriorates the performance slightly compared to using the non-adapted (zero-shot) source model. \newline
In contrast, the online adaptation works very well for both datasets. Compared to the benchmark, our results for the Dreyer2023 dataset are significantly better for every DA method except AdaBN ($p=0.125$).
For the Lee2019 dataset, all DA methods are significantly better than the benchmark method.

The unaveraged trial-wise accuracy is generally only slightly lower than the trial-wise accuracy, indicating that there are not many windows with an (over-) confident prediction. As expected, the window-wise accuracy is lower than the trial-wise accuracy.
\begin{table}
	\centering
	\caption{Dreyer2023 dataset cross-subject experiments. The stars after the method indicate the different significance levels ($p<0.05$ (*), $p<0.01$ (**) and $p<0.001$ (***)) compared to the benchmark method in the same setting.}
	\label{tab:results-large}
	\begin{tabular}{ |p{0.1cm}|p{2.8cm}||p{1.3cm}|p{1.3cm}|p{1.3cm}|  }
		\cline{2-5}
		\multicolumn{1}{c|}{}&Method  & TAcc(\%) & uTacc(\%) & WAcc(\%)\\
		\hline
		&BaseNet&$68.3\pm14.3$&$67.7\pm14.0$&$62.7\pm10.9$\\
		\hline
		\multirow{4}{*}{\rotatebox[origin=c]{90}{\begin{tabular}[c]{@{}c@{}}supervised\end{tabular}}}
		&RiemannMDM+PAR&$67.4\pm15.2$&$66.8\pm15.0$&$63.0\pm11.2$\\
		&BaseNet***&$70.8\pm15.1$&$70.0\pm14.9$&$64.1\pm11.4$\\ 
		&BaseNet+EA***&$72.5\pm15.6$&$71.9\pm15.4$&$66.1\pm12.1$\\
		&BaseNet+RA***&$72.6\pm15.5$&$72.1\pm15.3$&$66.1\pm12.2$\\
		\hline
		\multirow{6}{*}{\rotatebox[origin=c]{90}{\begin{tabular}[c]{@{}c@{}}unsupervised\end{tabular}}}
		&RiemannMDM&$67.4\pm15.2$&$66.7\pm14.9$&$63.0\pm11.2$\\ 
		&BaseNet+EA***&$70.0\pm15.0$&$69.6\pm14.6$&$63.8\pm11.1$\\
		&BaseNet+RA***&$70.1\pm15.1$&$69.6\pm14.8$&$63.9\pm11.3$\\
		&BaseNet+AdaBN**&$69.7\pm14.9$&$69.1\pm14.7$&$63.0\pm11.1$\\ 
		&BaseNet+EA+AdaBN***&$70.5\pm15.1$&$69.9\pm14.6$&$63.7\pm11.1$\\
		&BaseNet+RA+AdaBN***&$70.4\pm15.1$&$69.8\pm14.7$&$63.8\pm11.1$\\
		\hline
		\multirow{6}{*}{\rotatebox[origin=c]{90}{\begin{tabular}[c]{@{}c@{}}online\end{tabular}}}
		&RiemannMDM+GR&$69.3\pm14.8$&$68.6\pm14.3$&$63.5\pm11.3$\\
		&BaseNet+EA*&$70.7\pm15.0$&$70.2\pm14.8$&$64.0\pm11.2$\\
		&BaseNet+RA**&$70.9\pm15.0$&$70.4\pm14.6$&$64.2\pm11.3$\\
		&BaseNet+AdaBN&$70.2\pm15.0$&$69.1\pm14.5$&$62.7\pm10.9$\\ 
		&BaseNet+EA+AdaBN**&$71.2\pm14.7$&$70.3\pm14.2$&$63.5\pm10.6$\\
		&BaseNet+RA+AdaBN**&$71.2\pm14.7$&$70.4\pm14.2$&$63.5\pm10.7$\\
		
		\hline
	\end{tabular}
\end{table}

\begin{table}
	\centering
	\caption{Lee2019 cross-subject experiments. The stars after the method indicate the different significance levels ($p<0.05$ (*), $p<0.01$ (**) and $p<0.001$ (***)) compared to the benchmark method in the same setting.}
	\label{tab:results-openbmi}
	\begin{tabular}{ |p{0.1cm}|p{2.8cm}||p{1.3cm}|p{1.3cm}|p{1.3cm}|  }
		\cline{2-5}
		\multicolumn{1}{c|}{}&Method  & TAcc(\%) & uTacc(\%) & WAcc(\%)\\
		\hline
		&BaseNet&$72.4\pm14.7$&$71.6\pm14.5$&$66.4\pm12.0$\\
		\hline
		\multirow{4}{*}{\rotatebox[origin=c]{90}{\begin{tabular}[c]{@{}c@{}}supervised\end{tabular}}}
		&RiemannMDM+PAR&$65.1\pm12.7$&$64.7\pm13.0$&$62.6\pm10.5$\\
		&BaseNet***&$71.2\pm14.6$&$70.5\pm14.5$&$66.0\pm12.5$\\ 
		&BaseNet+EA***&$70.7\pm15.8$&$70.3\pm15.4$&$66.4\pm13.2$\\ 
		&BaseNet+RA***&$71.5\pm15.2$&$71.0\pm15.1$&$66.9\pm12.9$\\
		\hline
		\multirow{6}{*}{\rotatebox[origin=c]{90}{\begin{tabular}[c]{@{}c@{}}unsupervised\end{tabular}}}
		&RiemannMDM&$65.0\pm12.7$&$64.7\pm13.0$&$62.6\pm10.4$\\
		&BaseNet+EA***&$70.5\pm15.7$&$70.0\pm15.3$&$66.0\pm12.4$\\
		&BaseNet+RA***&$70.5\pm15.7$&$70.1\pm15.5$&$66.2\pm12.6$\\
		&BaseNet+AdaBN***&$71.6\pm14.1$&$70.9\pm14.0$&$66.1\pm12.0$\\
		&BaseNet+EA+AdaBN***&$70.2\pm15.2$&$69.7\pm14.8$&$65.8\pm12.3$\\
		&BaseNet+RA+AdaBN***&$70.3\pm15.2$&$70.1\pm15.0$&$66.1\pm12.4$\\
		\hline
		\multirow{6}{*}{\rotatebox[origin=c]{90}{\begin{tabular}[c]{@{}c@{}}online\end{tabular}}}
		&RiemannMDM+GR&$71.4\pm12.5$&$71.1\pm12.0$&$66.3\pm10.2$\\
		&BaseNet+EA***&$76.4\pm14.2$&$75.7\pm13.8$&$68.9\pm11.8$\\ 
		&BaseNet+RA***&$76.9\pm14.1$&$76.3\pm13.9$&$69.3\pm11.9$\\  
		&BaseNet+AdaBN*&$73.3\pm14.2$&$72.3\pm13.9$&$66.2\pm11.6$\\ 
		&BaseNet+EA+AdaBN***&$75.3\pm13.4$&$74.6\pm13.1$&$67.7\pm11.2$\\ 
		&BaseNet+RA+AdaBN***&$75.7\pm13.4$&$74.9\pm13.1$&$68.1\pm11.3$\\
		\hline
	\end{tabular}
\end{table}

% Outlook (Discussion) longer adaptation, joint DA?
\subsubsection{Results per window}
\begin{figure}
	\centering
	\begin{subfigure}[b]{0.5\textwidth}
		\centering
		\includegraphics[width=\columnwidth]{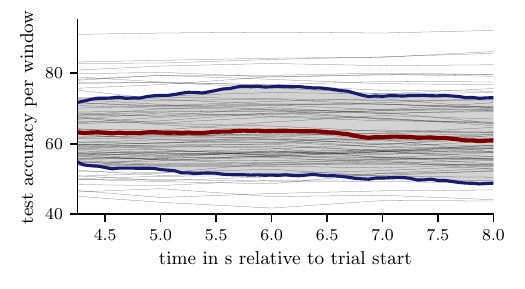}
		\caption{Dreyer2023 dataset}
		\label{fig:result_window-large}
	\end{subfigure}
	\hfill
	\begin{subfigure}[b]{0.5\textwidth}
		\centering
		\includegraphics[width=\columnwidth]{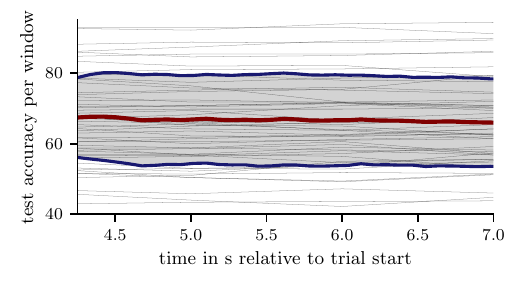}
		\caption{Lee2019 dataset}
		\label{fig:result_window-openbmi}
	\end{subfigure}
	\caption{Accuracies per window for BaseNet. Each gray line indicates one subject, the red line corresponds to the average and the blue lines correspond to the average $\pm$ standard deviation.}
	\label{fig:results-per-window}
\end{figure}

Another important aspect of online decoding is the performance throughout a trial to check whether a subject is able to perform the task long enough. Fig. \ref{fig:results-per-window} shows the window-wise accuracy over time. Generally, the accuracy is pretty stable over time, with only a slight decrease over time. This means that firstly, the subjects are able to perform the MI long enough and, secondly, our model does not have any boundary effects due to padding, cueing or joint decoding during training.
\subsubsection{Inference time}
\label{subsec:res-inference}
To validate the computational feasibility, we measured the inference time for a single window on a desktop CPU (Intel i7-1195G7 with 4 cores). The results showed that BaseNet had an inference time of \SI{2.15}{\milli\second}, while RiemannMDM achieved \SI{0.5}{\milli\second}. 
It is expected that DL models have a higher inference time than less complex traditional methods, especially on CPUs. Although BaseNets inference time is approximately four times higher than that of the benchmark method, both algorithms remain suitable for online decoding ($\ll\frac{1}{f_u}=\SI{62.5}{\milli\second}$).

\subsubsection{Spatial patterns}
\begin{figure}
	\centering
	\begin{subfigure}{0.5\textwidth}
		\centering
		\includegraphics[width=\columnwidth]{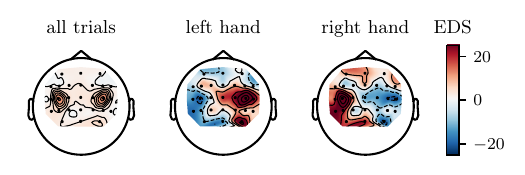}
		\caption{Dreyer2023 dataset}
		\label{fig:eds-large}
	\end{subfigure}
	
	\begin{subfigure}{0.5\textwidth}
		\centering
		\includegraphics[width=\columnwidth]{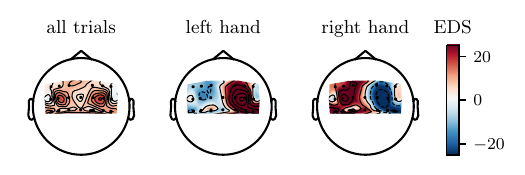}
		\caption{Lee2019 dataset}
		\label{fig:eds-openbmi}
	\end{subfigure}
	
	\caption{Per-class topoplots of the EDS scores for BaseNet and both datasets.}
	\label{fig:eds}
\end{figure}

To validate our results further, we employ the electrode discriminancy score (EDS) from \textit{Kumar et al.} \cite{kumar2024transfer}. The EDS calculates the difference between the original accuracy (of the source model) and the accuracy if one electrode is dropped (i.e., set to zero) during inference. Thus, a high EDS indicates a significant performance drop when that sensor is removed. The results are presented in Fig. \ref{fig:eds}. For both datasets, the C3 and C4 electrode yield the highest EDS. For the Lee2019 dataset other sensors such as the FC1 electrode and the sensors in the central parietal region of the cortex additionally exhibit high EDS. \newline
The class-wise EDS plots demonstrate that the contralateral sensorimotor regions consistently exhibit the highest EDS, specifically, the C4 electrode for left-hand motor imagery and the C3 electrode for right-hand motor imagery. These results are consistent with the neurophysiological principles of motor imagery,  where event-related desynchronization is predominantly observed in the contralateral sensorimotor areas during hand motor imagery \cite{pfurtscheller1999event, neuper2006erd}. \newline
Interestingly, negative EDS values are observed on the ipsilateral side of the motor cortex. This indicates that removing key features associated with right-hand trials improves accuracy for left-hand trials, and vice versa, as the removal reduces class confusion.

\section{Discussion}
The results show that our proposed method is an effective approach to tune existing DL models towards online MI decoding. In the following sections, we will discuss the prerequisites regarding data availability and data composition as well as the differences between the DA approaches and the general limitations of our investigation.
\subsection{Data availability and data composition}
Generally, data availability and composition dictate the overall training strategy. If there is insufficient data from other subjects to train a cross-subject decoder, a within-subject decoder using subject-specific data must be trained. In this case, a poor generalization of the decoder to other subjects is highly expected.\newline
Additionally, the cognitive and physical demands placed on the subject during data acquisition impact the volume of subject-specific data that can be collected. This, in turn, influences the selection of the appropriate decoding algorithm. For low amounts of data, traditional machine learning methods such as RiemannMDM from \cite{kumar2024transfer} or combinations of CSP and LDA are the better choice over DL solutions for within-subject BCI-MI decoding as evidenced by Table \ref{tab:results-within}. As more data becomes available, training within-subject DL models starts to get useful (compare Fig. \ref{fig:results-within}).

If there is enough data from other subjects, training in a cross-subject setting becomes feasible. This setting is advantageous for DL as the increased data volume typically enables DL models to exhibit better performances as shown in Table \ref{tab:results-large} and \ref{tab:results-openbmi}. The influence of the composition of data (i.e., how many subjects and how many runs/sessions per subject are available) is more complex. Generally, a high subject diversity improves the performance as evidenced in \cite{sartzetaki2023beyond} and our experiments on the Dreyer2023 dataset (compare Fig. \ref{fig:results-diversity-large}). For the Lee2019 dataset on the other hand, we observed that the composition of subjects and sessions had almost no influence on the final performance. We explain this by the domain shift introduced between offline and online session. Firstly, the sessions of the Lee2019 dataset were recorded on different days, and secondly, unlike the Dreyer2023 routine, there was no sham feedback during the offline sessions. These two aspects might have influenced the results in Fig. \ref{fig:results-diversity-openbmi}. The additional sessions introduce more data diversity compared to the additional data coming from the same session as in the Dreyer2023 dataset. Further, the domain shift between offline and online session in terms of recording procedure (i.e., visual feedback) \cite{shenoy2014combining} could likely be reduced through the inclusion of online sessions in the training data. 
%We added a small ablation in \ref{appendix-domain-shift} to further investigate this and found indications that the domain shift between offline and online is relevant for the Lee2019 dataset but without significance ($p=0.255$).

Besides the domain shift between offline and online data, there also exists an even bigger domain shift between the data from the source subjects and the data from the target subject due to the large subject-to-subject differences that are naturally present in EEG recordings. These domain shifts can be mitigated through the usage of different DA strategies depending on the availability of target data. 
\subsection{Domain adaptation}
The feasibility of certain DA strategies depends on the availability of target data. Using DL, supervised fine-tuning is the most effective strategy for the Dreyer2023 dataset, followed by online adaption. This effectiveness is likely due to the minimal domain shift between calibration and test data, enabling successful fine-tuning with offline data.
For the traditional ML approach on the other hand, the online adaptation strategy yields the best results followed by supervised and unsupervised fine-tuning, between which no differences are observed.

For the Lee2019 dataset, the same observations can be made for the traditional methods. However, the supervised and unsupervised fine-tuning approaches for BaseNet generally underperform the source performance. This can probably also be attributed to the shift between offline and online data which complicates fine-tuning with offline data. 
Notably, this dataset shows large improvements through online adaptation, further supporting our previous assumptions about the domain shift between the offline and online data. BaseNet using online alignment improves the initial source performance by over $4\%$ and RiemannMDM+GR outperforms RiemannMDM and RiemannMDM+PAR by over $6\%$.
Comparing the results from Table \ref{tab:results-large} and \ref{tab:results-openbmi}, the performance improvement of the online adaptation is larger for the Lee2019 dataset than for the Dreyer2023 dataset, i.e., the online adaptation helps to overcome the stronger inter-session domain shift in Lee2019.

Despite having the lowest data requirements regarding target data (calibration-free), the online adaptation methods yield excellent results for both datasets and both approaches (i.e., traditional methods and DL solutions). This supports the upcoming research of online adaptation methods for BCI MI decoding \cite{kumar2024transfer, li2023t, wimpff2024calibration}. Unexpected domain shifts will occur during inference, regardless of how well a BCI is designed (e.g., due to electrode movement, changes in the environment or changes in user behavior). Therefore, it is advisable to use online adaptation methods rather than offline adaptation methods. Additionally, online adaptation supports calibration-free BCI usage, allowing for instant user learning and thus potentially promoting an immediate process of continuous mutual learning between the user and the decoder. 
\subsection{Limitations}
The primary limitation of our study is that, although we investigated online adaptation, our experiments were conducted on previously recorded datasets, simulating a pseudo-online setting \cite{carrara2024pseudo}. Consequently, we did not examine the translation of model predictions into feedback signals \cite{stieger2021benefits}, which is another crucial aspect of closed-loop BCI systems. Both, the adaptation of the decoder during actual online use and the nature of the feedback signal can influence user behavior, potentially leading to further performance improvements. Despite these limitations, our pseudo-online design closely approximates real conditions, providing a solid foundation for future online studies.
%In other words, true closed-loop experiments are required in a future study to explore the mutual influence of online decoder adaptation and behavioral change of the user.
%\newline

Furthermore, the online adaptation time is fixed (i.e., the same number of online trials is available for all subjects) and relatively short. Consequently, we cannot accommodate user-specific needs, such as longer learning periods \cite{orsborn2014closed}. Due to the short training time per user (few trials, one session), our methods still need to be evaluated for their effectiveness concerning long-term changes within individual users. However, the adaptability demonstrated in this study offers promise that our proposed method could effectively handle such changes within users.
Additionally, longer learning periods would open up new possibilities for more sophisticated DA methods like mean teachers \cite{dobler2023robust} or certainty weighting \cite{marsden2024universal} that need more data and longer adaptation to become effective.%\newline
%Lastly, averaging the predictions of all windows within one trial only works for discrete MI paradigms (i.e., one task throughout the whole trial).
%%%%%%%%%%%%%%%%%%%%%%%%%%%%%%%%%%%%%%%%%%
\section{Conclusion}
In conclusion, our method RAP successfully adapts existing offline DL models for online decoding, overcoming the three primary challenges of employing DL for real-time BCI applications. By addressing the transition from offline to online models, reducing computational demands and minimizing subject-specific data requirements through source-free domain adaptation, our approach proves to be both powerful and potentially calibration-free.
Our experiments, conducted on a total of 133 subjects, reveal that online adaptation, despite having the lowest target data requirements, yields the best overall results. These findings demonstrate the potential of our method for practical real-time BCI applications and pave the way for developing co-adaptive, data-efficient DL-based BCI systems. 

\section*{Acknowledgment}
We would like to thank our colleagues and collaborators for their valuable input and support throughout this study. Additionally, we acknowledge the Quantum Human Machine Interfaces (QHMI) project (grant number 03ZU1110DC) within the QSens - Quantum Sensors of the Future Cluster for their financial support, which made this research possible.
\section*{Data Availability Statement}
The Dreyer2023 dataset \cite{dreyer2023large} is freely available at zenodo \url{https://zenodo.org/
records/8089820}. The Lee2019 dataset is freely available at GigaDB \url{http://gigadb.org/dataset/100542}.
\section*{Author Contributions}
Martin Wimpff: Conceptualization, methodology, software, validation, formal analysis, investigation, resources, data curation, writing—original draft preparation, writing—review and editing, visualization.
Jan Zerfowski.: Validation, writing—review and editing.
Bin Yang: Validation, writing—review and editing, supervision, project administration, funding acquisition.
\section*{Disclosure of interest}
The authors report there are no competing interests to declare.


%%%%%%%%%%%%%%%%%%%%%%%%%%%%%%%%%%%%%%%%%%%%%%%%%%%%%%%%%%%%%%%%%%%%%%%%%%%%%%%%

%%%%%%%%%%%%%%%%%%%%%%%%%%%%%%%%%%%%%%%%%%%%%%%%%%%%%%%%%%%%%%%%%%%%%%%%%%%%%%%%

%%%%%%%%%%%%%%%%%%%%%%%%%%%%%%%%%%%%%%%%%%%%%%%%%%%%%%%%%%%%%%%%%%%%%%%%%%%%%%%%
%\section*{APPENDIX}

%\subsection{Within-subject ablation}

%%%%%%%%%%%%%%%%%%%%%%%%%%%%%%%%%%%%%%%%%%%%%%%%%%%%%%%%%%%%%%%%%%%%%%%%%%%%%%%%


\begin{thebibliography}{99}
\bibitem{bcisociety}BCI Society, BCI Definition.  (2024), https://bcisociety.org/bci-definition/
\bibitem{peksa2023state}Peksa, J. \& Mamchur, D. State-of-the-Art on Brain-Computer Interface Technology. {\em Sensors}. \textbf{23}, 6001 (2023)
%\bibitem{bickenbach2011world}Bickenbach, J. The world report on disability. {\em Disability \& Society}. \textbf{26}, 655-658 (2011)
\bibitem{cervera2018brain}Cervera, M., Soekadar, S., Ushiba, J., Millán, J., Liu, M., Birbaumer, N. \& Garipelli, G. Brain-computer interfaces for post-stroke motor rehabilitation: a meta-analysis. {\em Annals Of Clinical And Translational Neurology}. \textbf{5}, 651-663 (2018)
\bibitem{soekadar2011erd}Soekadar, S., Witkowski, M., Mellinger, J., Ramos, A., Birbaumer, N. \& Cohen, L. ERD-based online brain-machine interfaces (BMI) in the context of neurorehabilitation: optimizing BMI learning and performance. {\em IEEE Transactions On Neural Systems And Rehabilitation Engineering}. \textbf{19}, 542-549 (2011)	
\bibitem{decety1996neurophysiological}Decety, J. The neurophysiological basis of motor imagery. {\em Behavioural Brain Research}. \textbf{77}, 45-52 (1996)
\bibitem{lee2019eeg}Lee, M., Kwon, O., Kim, Y., Kim, H., Lee, Y., Williamson, J., Fazli, S. \& Lee, S. EEG dataset and OpenBMI toolbox for three BCI paradigms: An investigation into BCI illiteracy. {\em GigaScience}. \textbf{8}, giz002 (2019)
\bibitem{sannelli2019large}Sannelli, C., Vidaurre, C., Müller, K. \& Blankertz, B. A large scale screening study with a SMR-based BCI: Categorization of BCI users and differences in their SMR activity. {\em PloS One}. \textbf{14}, e0207351 (2019)
\bibitem{zhang2020subject}Zhang, R., Li, F., Zhang, T., Yao, D. \& Xu, P. Subject inefficiency phenomenon of motor imagery brain-computer interface: Influence factors and potential solutions. {\em Brain Science Advances}. \textbf{6}, 224-241 (2020)
\bibitem{perdikis2018cybathlon}Perdikis, S., Tonin, L., Saeedi, S., Schneider, C. \& Millán, J. The Cybathlon BCI race: Successful longitudinal mutual learning with two tetraplegic users. {\em PLoS Biology}. \textbf{16}, e2003787 (2018)
\bibitem{korik2022competing}Korik, A., McCreadie, K., McShane, N., Du Bois, N., Khodadadzadeh, M., Stow, J., McElligott, J., Carroll, Á. \& Coyle, D. Competing at the Cybathlon championship for people with disabilities: long-term motor imagery brain–computer interface training of a cybathlete who has tetraplegia. {\em Journal Of NeuroEngineering And Rehabilitation}. \textbf{19}, 95 (2022)
\bibitem{mcfarland2018brain}McFarland, D. \& Wolpaw, J. Brain–computer interface use is a skill that user and system acquire together. {\em PLoS Biology}. \textbf{16}, e2006719 (2018)
\bibitem{shenoy2014combining}Shenoy, K. \& Carmena, J. Combining decoder design and neural adaptation in brain-machine interfaces. {\em Neuron}. \textbf{84}, 665-680 (2014)
\bibitem{orsborn2014closed}Orsborn, A., Moorman, H., Overduin, S., Shanechi, M., Dimitrov, D. \& Carmena, J. Closed-loop decoder adaptation shapes neural plasticity for skillful neuroprosthetic control. {\em Neuron}. \textbf{82}, 1380-1393 (2014)
\bibitem{sitaram2017closed}Sitaram, R., Ros, T., Stoeckel, L., Haller, S., Scharnowski, F., Lewis-Peacock, J., Weiskopf, N., Blefari, M., Rana, M., Oblak, E. \& Others Closed-loop brain training: the science of neurofeedback. {\em Nature Reviews Neuroscience}. \textbf{18}, 86-100 (2017)
\bibitem{kober2013learning}Kober, S., Witte, M., Ninaus, M., Neuper, C. \& Wood, G. Learning to modulate one's own brain activity: the effect of spontaneous mental strategies. {\em Frontiers In Human Neuroscience}. \textbf{7} pp. 695 (2013)
\bibitem{gaume2016psychoengineering}Gaume, A., Vialatte, A., Mora-Sánchez, A., Ramdani, C. \& Vialatte, F. A psychoengineering paradigm for the neurocognitive mechanisms of biofeedback and neurofeedback. {\em Neuroscience \& Biobehavioral Reviews}. \textbf{68} pp. 891-910 (2016)
\bibitem{mladenovic2018generic}Mladenović, J., Mattout, J. \& Lotte, F. A generic framework for adaptive EEG-based BCI training and operation. {\em Brain–Computer Interfaces Handbook}. pp. 595-612 (2018)
\bibitem{craik2019deep}Craik, A., He, Y. \& Contreras-Vidal, J. Deep learning for electroencephalogram (EEG) classification tasks: a review. {\em Journal Of Neural Engineering}. \textbf{16}, 031001 (2019)
\bibitem{vavoulis2023review}Vavoulis, A., Figueiredo, P. \& Vourvopoulos, A. A Review of Online Classification Performance in Motor Imagery-Based Brain-Computer Interfaces for Stroke Neurorehabilitation. {\em Signals}. \textbf{4}, 73-86 (2023)
\bibitem{tayeb2019validating}Tayeb, Z., Fedjaev, J., Ghaboosi, N., Richter, C., Everding, L., Qu, X., Wu, Y., Cheng, G. \& Conradt, J. Validating deep neural networks for online decoding of motor imagery movements from EEG signals. {\em Sensors}. \textbf{19}, 210 (2019)
\bibitem{jeong2020brain}Jeong, J., Shim, K., Kim, D. \& Lee, S. Brain-controlled robotic arm system based on multi-directional CNN-BiLSTM network using EEG signals. {\em IEEE Transactions On Neural Systems And Rehabilitation Engineering}. \textbf{28}, 1226-1238 (2020)
\bibitem{karacsony2019brain}Karácsony, T., Hansen, J., Iversen, H. \& Puthusserypady, S. Brain computer interface for neuro-rehabilitation with deep learning classification and virtual reality feedback. {\em Proceedings Of The 10th Augmented Human International Conference 2019}. pp. 1-8 (2019)
\bibitem{stieger2021benefits}Stieger, J., Engel, S., Suma, D. \& He, B. Benefits of deep learning classification of continuous noninvasive brain–computer interface control. {\em Journal Of Neural Engineering}. \textbf{18}, 046082 (2021)
\bibitem{forenzo2024continuous}Forenzo, D., Zhu, H., Shanahan, J., Lim, J. \& He, B. Continuous tracking using deep learning-based decoding for noninvasive brain–computer interface. {\em PNAS Nexus}. \textbf{3}, pgae145 (2024)
%\bibitem{wimpff2024towards}Wimpff, M., Zerfowski, J. \& Yang, B. Towards calibration-free online EEG motor imagery decoding using Deep Learning. \textit{ESANN 2024 Proceedings} (2024), \textit{accepted}.
\bibitem{lawhern2018eegnet}Lawhern, V., Solon, A., Waytowich, N., Gordon, S., Hung, C. \& Lance, B. EEGNet: a compact convolutional neural network for EEG-based brain-computer interfaces. {\em Journal Of Neural Engineering}. \textbf{15}, 056013 (2018)
\bibitem{kumar2024transfer}Kumar, S., Alawieh, H., Racz, F., Fakhreddine, R. \& Millán, J. Transfer learning promotes acquisition of individual BCI skills. {\em PNAS Nexus}. \textbf{3}, pgae076 (2024)
\bibitem{dreyer2023large}Dreyer, P., Roc, A., Pillette, L., Rimbert, S. \& Lotte, F. A large EEG database with users’ profile information for motor imagery brain-computer interface research. {\em Scientific Data}. \textbf{10}, 580 (2023)
\bibitem{sartzetaki2023beyond}Sartzetaki, C., Antoniadis, P., Antonopoulos, N., Gkinis, I., Krasoulis, A., Perdikis, S. \& Pitsikalis, V. Beyond Within-Subject Performance: A Multi-Dataset Study of Fine-Tuning in the EEG Domain. {\em 2023 IEEE International Conference On Systems, Man, And Cybernetics (SMC)}. pp. 4429-4435 (2023)
\bibitem{wimpff2024eeg}Wimpff, M., Gizzi, L., Zerfowski, J. \& Yang, B. EEG motor imagery decoding: A framework for comparative analysis with channel attention mechanisms. {\em Journal Of Neural Engineering}. (2024)
\bibitem{schirrmeister2017deep}Schirrmeister, R., Springenberg, J., Fiederer, L., Glasstetter, M., Eggensperger, K., Tangermann, M., Hutter, F., Burgard, W. \& Ball, T. Deep learning with convolutional neural networks for EEG decoding and visualization. {\em Human Brain Mapping}. \textbf{38}, 5391-5420 (2017)
\bibitem{wu2022transfer}Wu, D., Jiang, X. \& Peng, R. Transfer learning for motor imagery based brain–computer interfaces: A tutorial. {\em Neural Networks}. \textbf{153} pp. 235-253 (2022)
\bibitem{ko2021survey}Ko, W., Jeon, E., Jeong, S., Phyo, J. \& Suk, H. A survey on deep learning-based short/zero-calibration approaches for EEG-based brain–computer interfaces. {\em Frontiers In Human Neuroscience}. \textbf{15} pp. 643386 (2021)
\bibitem{kostas2020thinker}Kostas, D. \& Rudzicz, F. Thinker invariance: enabling deep neural networks for BCI across more people. {\em Journal Of Neural Engineering}. \textbf{17}, 056008 (2020)
\bibitem{sultana2023towards}Sultana, M., Reichert, C., Sweeney-Reed, C. \& Perdikis, S. Towards Calibration-Less BCI-Based Rehabilitation. {\em 2023 IEEE International Conference On Metrology For EXtended Reality, Artificial Intelligence And Neural Engineering (MetroXRAINE)}. pp. 11-16 (2023)
\bibitem{gu2023generalizable}Gu, X., Han, J., Yang, G. \& Lo, B. Generalizable Movement Intention Recognition with Multiple Heterogeneous EEG Datasets. {\em 2023 IEEE International Conference On Robotics And Automation (ICRA)}. pp. 9858-9864 (2023)
\bibitem{han2023eeg}Han, J., Wei, X. \& Faisal, A. EEG decoding for datasets with heterogenous electrode configurations using transfer learning graph neural networks. {\em Journal Of Neural Engineering}. \textbf{20}, 066027 (2023)
\bibitem{he2020different}He, H. \& Wu, D. Different set domain adaptation for brain-computer interfaces: A label alignment approach. {\em IEEE Transactions On Neural Systems And Rehabilitation Engineering}. \textbf{28}, 1091-1108 (2020)
\bibitem{ju2020federated}Ju, C., Gao, D., Mane, R., Tan, B., Liu, Y. \& Guan, C. Federated transfer learning for EEG signal classification. {\em 2020 42nd Annual International Conference Of The IEEE Engineering In Medicine \& Biology Society (EMBC)}. pp. 3040-3045 (2020)
\bibitem{wang2023privacy}Wang, K., Yang, M., Li, C., Liu, A., Qian, R. \& Chen, X. Privacy-Preserving Domain Adaptation for Intracranial EEG Classification via Information Maximization and Gaussian Mixture Model. {\em IEEE Sensors Journal}. (2023)
\bibitem{guetschel2023transfer}Guetschel, P. \& Tangermann, M. Transfer Learning between Motor Imagery Datasets using Deep Learning–Validation of Framework and Comparison of Datasets. {\em ArXiv Preprint ArXiv:2311.16109}. (2023)
\bibitem{xie2023cross}Xie, Y., Wang, K., Meng, J., Yue, J., Meng, L., Yi, W., Jung, T., Xu, M. \& Ming, D. Cross-dataset transfer learning for motor imagery signal classification via multi-task learning and pre-training. {\em Journal Of Neural Engineering}. \textbf{20}, 056037 (2023)
\bibitem{ouahidi2023strong}Ouahidi, Y., Gripon, V., Pasdeloup, B., Bouallegue, G., Farrugia, N. \& Lioi, G. A Strong and Simple Deep Learning Baseline for BCI MI Decoding. {\em ArXiv Preprint ArXiv:2309.07159}. (2023)
\bibitem{junqueira2024systematic}Junqueira, B., Aristimunha, B., Chevallier, S. \& Camargo, R. A systematic evaluation of Euclidean alignment with deep learning for EEG decoding. {\em Journal Of Neural Engineering}. (2024)
\bibitem{duan2021ultra}Duan, T., Chauhan, M., Shaikh, M., Chu, J. \& Srihari, S. Ultra Efficient Transfer Learning with Meta Update for Continuous EEG Classification Across Subjects.. {\em Canadian Conference On AI}. (2021)
\bibitem{liu2022subject}Liu, S., Zhang, J., Wang, A., Wu, H., Zhao, Q. \& Long, J. Subject adaptation convolutional neural network for EEG-based motor imagery classification. {\em Journal Of Neural Engineering}. \textbf{19}, 066003 (2022)
\bibitem{xu2021selective}Xu, Y., Huang, X. \& Lan, Q. Selective cross-subject transfer learning based on riemannian tangent space for motor imagery brain-computer interface. {\em Frontiers In Neuroscience}. \textbf{15} pp. 779231 (2021)
\bibitem{an2020few}An, S., Kim, S., Chikontwe, P. \& Park, S. Few-shot relation learning with attention for EEG-based motor imagery classification. {\em 2020 IEEE/RSJ International Conference On Intelligent Robots And Systems (IROS)}. pp. 10933-10938 (2020)
\bibitem{ouahidi2024unsupervised}Ouahidi, Y., Lioi, G., Farrugia, N., Pasdeloup, B. \& Gripon, V. Unsupervised Adaptive Deep Learning Method For BCI Motor Imagery Decoding. {\em ArXiv Preprint ArXiv:2403.15438}. (2024)
\bibitem{xu2021improving}Xu, L., Ma, Z., Meng, J., Xu, M., Jung, T. \& Ming, D. Improving transfer performance of deep learning with adaptive batch normalization for brain-computer interfaces. {\em 2021 43rd Annual International Conference Of The IEEE Engineering In Medicine \& Biology Society (EMBC)}. pp. 5800-5803 (2021)
\bibitem{zoumpourlis2024motor}Zoumpourlis, G. \& Patras, I. Motor imagery decoding using ensemble curriculum learning and collaborative training. {\em 2024 12th International Winter Conference On Brain-Computer Interface (BCI)}. pp. 1-8 (2024)
\bibitem{bakas2023latent}Bakas, S., Ludwig, S., Adamos, D., Laskaris, N., Panagakis, Y. \& Zafeiriou, S. Latent Alignment with Deep Set EEG Decoders. {\em ArXiv Preprint ArXiv:2311.17968}. (2023)
\bibitem{xia2022privacy}Xia, K., Deng, L., Duch, W. \& Wu, D. Privacy-preserving domain adaptation for motor imagery-based brain-computer interfaces. {\em IEEE Transactions On Biomedical Engineering}. \textbf{69}, 3365-3376 (2022)
\bibitem{zhuo2024riemannian}Zhuo, F., Zhang, X., Tang, F., Yu, Y. \& Liu, L. Riemannian transfer learning based on log-Euclidean metric for EEG classification. {\em Frontiers In Neuroscience}. \textbf{18} pp. 1381572 (2024)
\bibitem{jimenez2020custom}Jiménez-Guarneros, M. \& Gómez-Gil, P. Custom Domain Adaptation: A new method for cross-subject, EEG-based cognitive load recognition. {\em IEEE Signal Processing Letters}. \textbf{27} pp. 750-754 (2020)
\bibitem{zhang2020manifold}Zhang, W. \& Wu, D. Manifold embedded knowledge transfer for brain-computer interfaces. {\em IEEE Transactions On Neural Systems And Rehabilitation Engineering}. \textbf{28}, 1117-1127 (2020)
\bibitem{mao2023online}Mao, T., Li, C., Zhao, Y., Song, R. \& Chen, X. Online test-time adaptation for patient-independent seizure prediction. {\em IEEE Sensors Journal}. (2023)
\bibitem{li2023t}Li, S., Wang, Z., Luo, H., Ding, L. \& Wu, D. T-TIME: Test-time information maximization ensemble for plug-and-play BCIs. {\em IEEE Transactions On Biomedical Engineering}. (2023)
\bibitem{wimpff2024calibration}Wimpff, M., Döbler, M. \& Yang, B. Calibration-free online test-time adaptation for electroencephalography motor imagery decoding. {\em 2024 12th International Winter Conference On Brain-Computer Interface (BCI)}. pp. 1-6 (2024)
\bibitem{he2019transfer}He, H. \& Wu, D. Transfer learning for brain–computer interfaces: A Euclidean space data alignment approach. {\em IEEE Transactions On Biomedical Engineering}. \textbf{67}, 399-410 (2019)
\bibitem{zanini2017transfer}Zanini, P., Congedo, M., Jutten, C., Said, S. \& Berthoumieu, Y. Transfer learning: A Riemannian geometry framework with applications to brain–computer interfaces. {\em IEEE Transactions On Biomedical Engineering}. \textbf{65}, 1107-1116 (2017)
\bibitem{xu2020cross}Xu, L., Xu, M., Ke, Y., An, X., Liu, S. \& Ming, D. Cross-dataset variability problem in EEG decoding with deep learning. {\em Frontiers In Human Neuroscience}. \textbf{14} pp. 103 (2020)
\bibitem{li2016revisiting}Li, Y., Wang, N., Shi, J., Liu, J. \& Hou, X. Revisiting batch normalization for practical domain adaptation. {\em ArXiv Preprint ArXiv:1603.04779}. (2016)
\bibitem{schneider2020improving}Schneider, S., Rusak, E., Eck, L., Bringmann, O., Brendel, W. \& Bethge, M. Improving robustness against common corruptions by covariate shift adaptation. {\em Advances In Neural Information Processing Systems}. \textbf{33} pp. 11539-11551 (2020)
\bibitem{carrara2024pseudo}Carrara, I. \& Papadopoulo, T. Pseudo-online framework for BCI evaluation: a MOABB perspective using various MI and SSVEP datasets. {\em Journal Of Neural Engineering}. \textbf{21}, 016003 (2024)
\bibitem{dobler2023robust}Döbler, M., Marsden, R. \& Yang, B. Robust mean teacher for continual and gradual test-time adaptation. {\em Proceedings Of The IEEE/CVF Conference On Computer Vision And Pattern Recognition}. pp. 7704-7714 (2023)
\bibitem{marsden2024universal}Marsden, R., Döbler, M. \& Yang, B. Universal test-time adaptation through weight ensembling, diversity weighting, and prior correction. {\em Proceedings Of The IEEE/CVF Winter Conference On Applications Of Computer Vision}. pp. 2555-2565 (2024)
\bibitem{pfurtscheller1999event}Pfurtscheller, G. \& Da Silva, F. Event-related EEG/MEG synchronization and desynchronization: basic principles. {\em Clinical Neurophysiology}. \textbf{110}, 1842-1857 (1999)
\bibitem{neuper2006erd}Neuper, C., Wörtz, M. \& Pfurtscheller, G. ERD/ERS patterns reflecting sensorimotor activation and deactivation. {\em Progress In Brain Research}. \textbf{159} pp. 211-222 (2006)

\end{thebibliography}
\end{document}